\documentclass[sigconf]{acmart}

\AtBeginDocument{%
  }

\setcopyright{acmlicensed}
\copyrightyear{2018}
\acmYear{2018}
\acmDOI{XXXXXXX.XXXXXXX}

\acmConference[Conference acronym 'XX]{Make sure to enter the correct
  conference title from your rights confirmation emai}{June 03--05,
  2018}{Woodstock, NY}
\acmISBN{978-1-4503-XXXX-X/18/06}

\begin{document}

\title{From Complexity to Simplicity: Using Python Instead of PsychoPy for fNIRS Data Collection}

\author{Shayla Sharmin}

\orcid{0000-0001-5137-1301}

\affiliation{%
  \institution{University of Delaware}
  \city{Newark}
  \state{Delaware}
  \country{USA}
  \postcode{19716}
}
\author{Md. Fahim Abrar}

\orcid{0000-0001-5137-1301}
\affiliation { \institution{University of Delaware}
  \city{Newark}
  \state{Delaware}
  \country{USA}
  \postcode{19716}
}
\author{Roghayeh Leila Barmaki}
\affiliation{%
  \institution{University of Delaware}
  \city{Newark}
  \state{DE}
  \country{U.S.A}}
\orcid{0000-0002-7570-5270}

\renewcommand{\shortauthors}{Sharmin et al.}

\begin{abstract}
 Functional near-infrared spectroscopy (fNIRS) is a non-invasive optical technique that measures brain activity by estimating blood oxygenation using near-infrared light. Traditionally, PsychoPy is used in many studies to send task-specific markers, requiring a separate device to interface with the fNIRS data collection system. In this work, we present a Python-based implementation to send markers directly, eliminating the need for an additional device. This approach allows researchers to run both marker transmission and fNIRS data collection on the same computer, simplifying the setup and enhancing accessibility. This streamlined solution reduces hardware requirements and makes fNIRS studies more efficient. 
\end{abstract}


\begin{teaserfigure}
  \includegraphics[width=\textwidth]{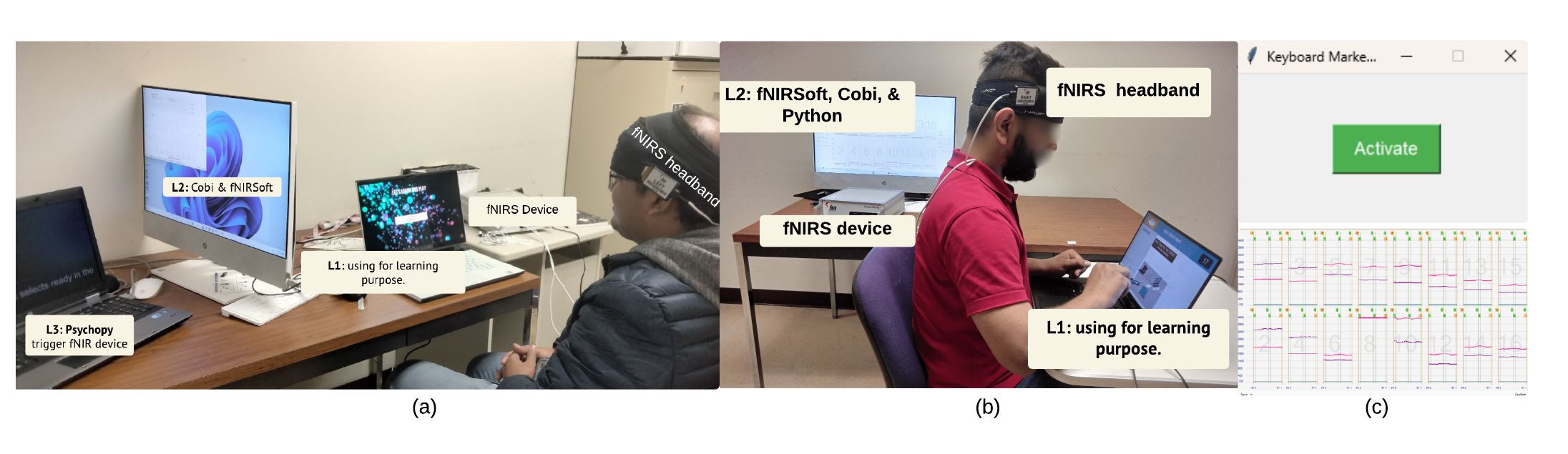}
  \caption{(a) original setup: additional laptop to use Psychopy (b) improved setup: no additional laptop for python (c) GUI for Python code (up) and fNIRS data with marker}
  \Description{}
  \label{fig:teaser}
\end{teaserfigure}

\received{20 February 2007}
\received[revised]{12 March 2009}
\received[accepted]{5 June 2009}

\maketitle

\section{Introduction}
Basic code can greatly improve the usability of an Human Computer Interaction study by simplifying the comprehension, modification, and maintenance of the experimental setup for researchers. Code that is concise and easily understood decreases the probability of errors and simplifies the process of debugging, guaranteeing a smooth execution of the study. Easy-to-understand coding simplifies making changes and adjustments depending on participant feedback, resulting in more streamlined and successful study outputs.
 Streamlining the code base also facilitates the integration of new team members, enabling them to make meaningful contributions to the project with minimal time needed to get up to speed.
In our study we compared game based learning and video based learning using fNIRS to measure brain activity that allows us to understand the cognitive processes involved during the tasks. Originally, our setup required three laptops, including one running a complex PsychoPy script to send biomarkers to the fNIRS data collection software fNIRSoft. In these experiments we have used fNIRS imager 1200/2000S
(fNIRS Device LLC, Potomac,
MD, USA) by Biopac \cite{AYAZ201236} that comes with COBI and fNIRSoft software. COBI is used to collect fNIRS data and fNIRSoft is used to pre-process, analysis, and visualize the data. 
Python's versatility, ease of use, and rich libraries make it useful in physiological research. It excels at data acquisition and processing with PyEPL, BioSPPy, and NeuroKit for device interfaces (ECG, EEG, EMG) and NumPy, SciPy, and Pandas for numerical computing and data manipulation. 
PsychoPy, an open-source software package, is used in psychology and neuroscience research to create and perform experiments, surveys, and behavioral studies by collecting and recording data. 
PsychoPy has precise accuracy for reaction time and stimulus duration measurements, cross-platform support on Windows, macOS, and Linux, and many built-in routines for presenting varied stimuli. It enables complicated and customized Python scripted experiments, has extensive data gathering and analysis tools, and is developed and improved by a large, active open-source community. 

This paper discusses the improvements made by replacing the PsychoPy script with a simplified Python code, reducing the number of devices and enhancing the experiment's overall efficiency.

\section{Development of the Experimental Setup}
\subsection{Original Setup}
In the original setup, three laptops were used (\autoref{fig:teaser}(a)) \cite{10445542}:
\textbf{L1:} a laptop running the learning materials.
\textbf{L2:} Desktop computer with an Intel(R) Core(TM) i7-10700T CPU at 2.00 GHz, connected to the fNIRS device, running Cognitive Optical Brain Imaging Studio Software (COBI) for data collection and fNIRSoft Software (Version 4.9) for analysis.
\textbf{L3:} Laptop with an Intel(R) Core(TM) i5 CPU M 460 @ 2.53 GHz, connected to L2 via a serial port, running a PsychoPy script for stimulus presentation and triggering the fNIRS device.
\subsection{Revised Setup}
The revised setup simplifies the experimental process (\autoref{fig:teaser}(b)) \cite{sharmin2024fnirs}:
\textbf{Reduced Equipment:} Only L1 and L2 are used, eliminating the need for L3.
\textbf{Python Code:} A Python script replaces the PsychoPy script for sending markers and dividing tasks. First a GUI appeared on the screen ((\autoref{fig:teaser}(c) (up)). Then this script uses the keyboard to send signals to L2, marking different blocks of the experiment such as concept definition, rest, quiz, and feedback in fNIRS data ((\autoref{fig:teaser}(c) (down)).
\section{Preliminary Evaluation}
\begin{figure} [h]
\centering
  \includegraphics[width=0.7\linewidth]{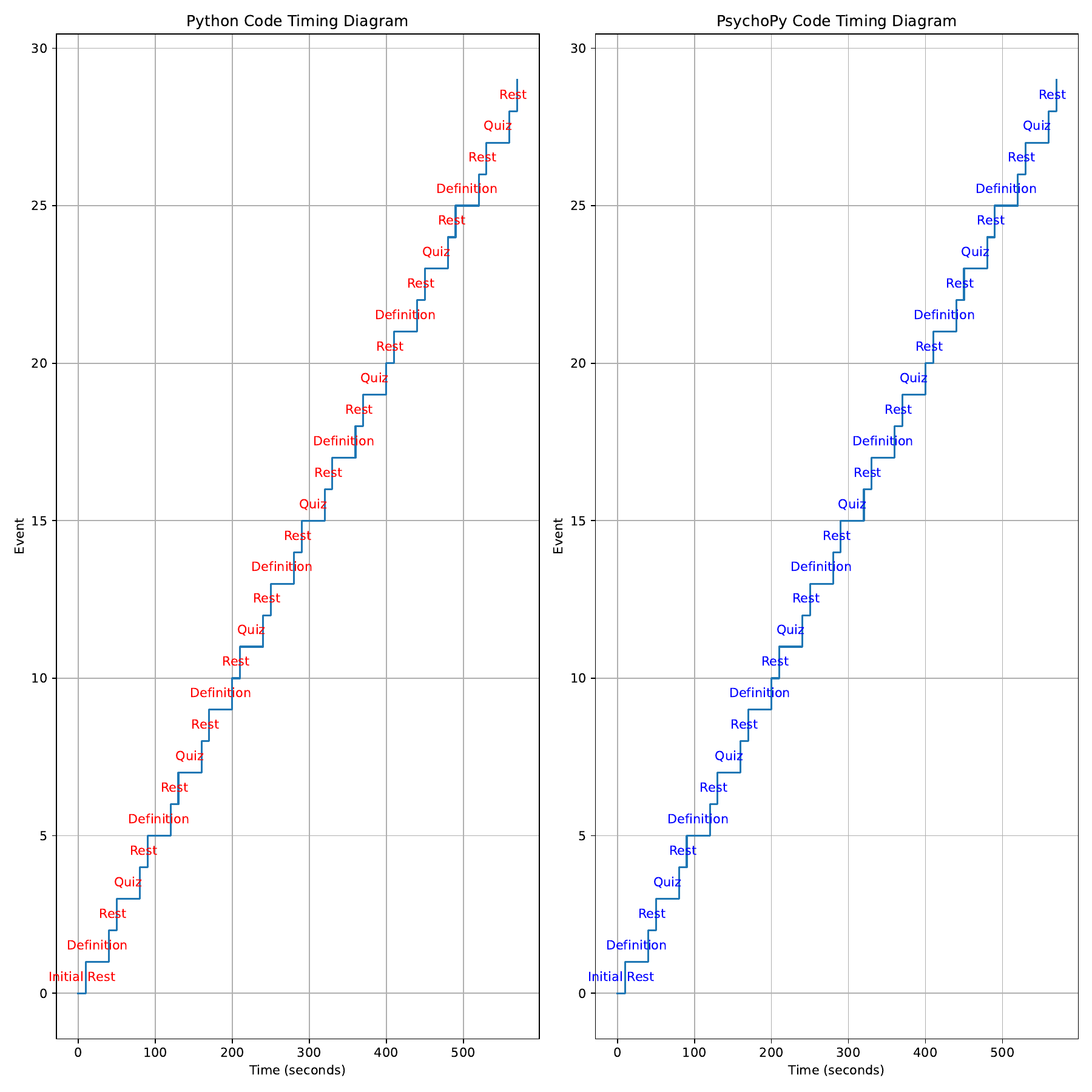}
  \caption{ Python and Psychopy Timing figure}
  \label{fig:timingFig}
\end{figure}

The Python code uses sequential ``time.sleep'' calls, while PsychoPy uses its routine management system for precise timing. The generated timing diagrams confirm that both the Python and PsychoPy implementations follow the same sequence of rest and quiz periods with identical timing intervals.Both diagrams correctly accumulate the time intervals, showing a linear progression of time with steps corresponding to each event's duration. This visual comparison highlights that the timing and sequence of events are the same (see figure \ref{fig:timingFig}).

\subsection{Complexity Analysis}
\paragraph{\textbf{Time and Space Complexity}}
 Both the Python code and the PsychoPy XML code have a linear time complexity of O(n), where n represents the number of loops, as they execute actions in a sequential manner. The space complexity for both algorithms is O(1) since they use a constant amount of memory, regardless of the number of loops or references to predefined routines.

\subsection{ Benefits of the Revised Setup}
The revised Python Code is written in Python and the
PsychoPy Code is written in XML, a markup language designed to store and transport data.
 Python is more intuitive for defining logic and functionality compared to XML, which is verbose and less suited for complex logic.
\begin{itemize}
\item[]
\item \textbf{Reduced Complexity:} The revised code with Python introduces fewer devices and simpler code by replacing the complex PsychoPy code that streamline the experiment setup, reducing potential points of failure.

\item[] \textbf{Simplified Coding:} Python code is more easily comprehensible and adaptable, enhancing the experiment's versatility.Utilizes commonly used Python libraries (pyautogui, tkinter, threading) that are both user-friendly and widely used.
The Python code consists of roughly 50 lines, while the PsychoPy code consists of several hundred lines.
 The Python code is notably shorter and more clear, simplifying comprehension and maintenance.

\item \textbf{Integration and GUI:} The revised code is easily integrates with other Python scripts and applications.
It also provides a graphical user interface for starting the task, making it user-friendly. Python code, especially when using pyautogui and tkinter, is straightforward and readable, making it accessible to a broader range of developers.
The GUI created with ``tkinter'' is simple and can be easily modified to add more features.
\item \textbf{Debugging}: It is easier to debug with standard Python tools. Debugging Python scripts is more straightforward with tools like PythonDebugger (PDB) and interactive environments like Jupyter Notebooks.
Testing Python code can be easily done using unit testing frameworks like unittest or pytest.
\item \textbf{Error Handling} Python's try-except blocks allow for robust error handling, which can be crucial in managing unexpected scenarios during runtime.

\end {itemize}

\section{Conclusion}
While both implementations achieve the same functionality, the Python script offers several advantages for HCI experiments.
Python is not confined to a specific framework, allowing for seamless integration with other tools and libraries. Python’s concise and readable syntax makes it easier to understand, modify, and extend.
The extensive Python community provides ample resources, ensuring that researchers can find support and solutions to problems quickly. Where PsychoPy support is smaller, specialized community focused on psychological research.
For researchers in HCI, Python presents a more versatile and user-friendly option for experiment development, making it a superior choice over specialized frameworks like PsychoPy in many scenarios.

\bibliographystyle{ACM-Reference-Format}
\bibliography{reference}


\begin{thebibliography}{3}


\ifx \showCODEN    \undefined \def \showCODEN     #1{\unskip}     \fi
\ifx \showDOI      \undefined \def \showDOI       #1{#1}\fi
\ifx \showISBNx    \undefined \def \showISBNx     #1{\unskip}     \fi
\ifx \showISBNxiii \undefined \def \showISBNxiii  #1{\unskip}     \fi
\ifx \showISSN     \undefined \def \showISSN      #1{\unskip}     \fi
\ifx \showLCCN     \undefined \def \showLCCN      #1{\unskip}     \fi
\ifx \shownote     \undefined \def \shownote      #1{#1}          \fi
\ifx \showarticletitle \undefined \def \showarticletitle #1{#1}   \fi
\ifx \showURL      \undefined \def \showURL       {\relax}        \fi
\providecommand\bibfield[2]{#2}
\providecommand\bibinfo[2]{#2}
\providecommand\natexlab[1]{#1}
\providecommand\showeprint[2][]{arXiv:#2}

\bibitem[Ayaz et~al\mbox{.}(2012)]%
        {AYAZ201236}
\bibfield{author}{\bibinfo{person}{Hasan Ayaz}, \bibinfo{person}{Patricia~A. Shewokis}, \bibinfo{person}{Scott Bunce}, \bibinfo{person}{Kurtulus Izzetoglu}, \bibinfo{person}{Ben Willems}, {and} \bibinfo{person}{Banu Onaral}.} \bibinfo{year}{2012}\natexlab{}.
\newblock \showarticletitle{Optical brain monitoring for operator training and mental workload assessment}.
\newblock \bibinfo{journal}{\emph{NeuroImage}} \bibinfo{volume}{59}, \bibinfo{number}{1} (\bibinfo{year}{2012}), \bibinfo{pages}{36--47}.
\newblock
\showISSN{1053-8119}
\urldef\tempurl%
\url{https://doi.org/10.1016/j.neuroimage.2011.06.023}
\showDOI{\tempurl}
\newblock
\shownote{Neuroergonomics: The human brain in action and at work}.


\bibitem[Sharmin et~al\mbox{.}(2024a)]%
        {sharmin2024fnirs}
\bibfield{author}{\bibinfo{person}{Shayla Sharmin}, \bibinfo{person}{Elham Bakhshipour}, \bibinfo{person}{Behdokht Kiafar}, \bibinfo{person}{Md~Fahim Abrar}, \bibinfo{person}{Pinar Kullu}, \bibinfo{person}{Nancy Getchell}, {and} \bibinfo{person}{Roghayeh~Leila Barmaki}.} \bibinfo{year}{2024}\natexlab{a}.
\newblock \showarticletitle{{fNIRS Analysis of Interaction Techniques in Touchscreen-Based Educational Gaming}}.
\newblock \bibinfo{journal}{\emph{arXiv preprint arXiv:2405.08906}} (\bibinfo{date}{May} \bibinfo{year}{2024}).
\newblock


\bibitem[Sharmin et~al\mbox{.}(2024b)]%
        {10445542}
\bibfield{author}{\bibinfo{person}{Shayla Sharmin}, \bibinfo{person}{Reza Koiler}, \bibinfo{person}{Rifat Sadik}, \bibinfo{person}{Arpan Bhattacharjee}, \bibinfo{person}{Priyanka~Raju Patre}, \bibinfo{person}{Pinar Kullu}, \bibinfo{person}{Charles Hohensee}, \bibinfo{person}{Nancy Getchell}, {and} \bibinfo{person}{Roghayeh~Leila Barmaki}.} \bibinfo{year}{2024}\natexlab{b}.
\newblock \showarticletitle{Cognitive Engagement for STEM+C Education: Investigating Serious Game Impact on Graph Structure Learning with fNIRS}. In \bibinfo{booktitle}{\emph{2024 IEEE International Conference on Artificial Intelligence and eXtended and Virtual Reality (AIxVR)}}. \bibinfo{pages}{195--204}.
\newblock
\urldef\tempurl%
\url{https://doi.org/10.1109/AIxVR59861.2024.00032}
\showDOI{\tempurl}


\end{thebibliography}

\end{document}